\documentclass[prb,preprint,amsmath,amssymb]{revtex4}
\usepackage{color}
\begin{document}
\author{Kevin Leung,$^{1*}$ Noah B.~Schorr,$^1$ Matthew Mayer,$^2$
Timothy N.~Lambert,$^1$ Y.~Shirley Meng,$^2$ and Katharine L.~Harrison$^1$}
\affiliation{$^1$Sandia National Laboratories, 
Albuquerque, NM 87185, United States\\
{\tt kleung@sandia.gov}\\
$^2$NanoEngineering, University of California at San Diego\\
}
\date{\today}
\title{Edge-Propagation Discharge Mechanism in CF$_x$ Batteries
-- a First Principles and Experimental Study}

\input epsf.sty
%\ssp
\renewcommand{\thetable}{\arabic{table}}

\begin{abstract}

Graphite fluoride (CF$_x$) cathodes coupled with lithium anodes yield one
of the highest theoretical specific capacities ($>$860 mAh/g) among primary
batteries.  In practice, the observed discharge voltage ($\sim$2.5~V)
is significantly lower than thermodynamic limits ($>$4.5~V), the discharge
rate is low, and so far Li/CF$_x$ has only been used in primary batteries.
Understanding the discharge mechanism at atomic length scales will improve
practical CF$_x$ energy density, rate capability, and rechargeability.  So
far, purely experimental techniques have not identified the correct discharge
mechanism or explained the discharge voltage.  We apply Density Functional
Theory calculations to demonstrate that a CF$_x$-edge propagation discharge
mechanism based on lithium insertion at the CF/C boundary in partially
discharged CF$_x$ exhibits a voltage range of 2.5 to 2.9~V --- depending on
whether solvent molecules are involved.  The voltages and solvent dependence
agrees with our discharge and galvanostatic intermittent titration technique
measurements.  The predicted
discharge kinetics are consistent with CF$_x$ operations.  Finally, we
predict Li/CF$_x$ rechargeability under the application of high potentials,
along a charging pathway different from that of discharge.  Our work
represents a general, quasi-kinetic framework to understand the discharge
of conversion cathodes, circumventing the widely used phase diagram approach
which most likely does not apply to Li/CF$_x$ because equilibrium conditions
are not attained in this system.

\end{abstract}

\maketitle

\section{Introduction}
 
\begin{figure}
\centerline{\hbox{ (a) \epsfxsize=2.00in \epsfbox{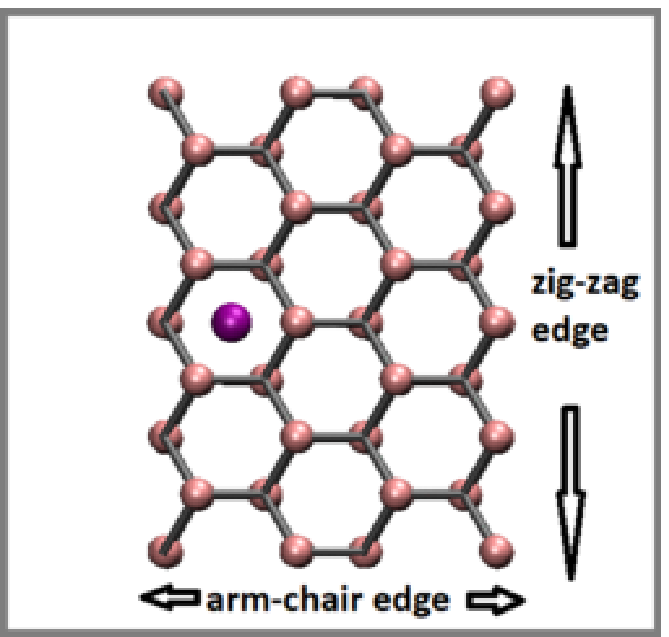} 
		   \epsfxsize=2.00in \epsfbox{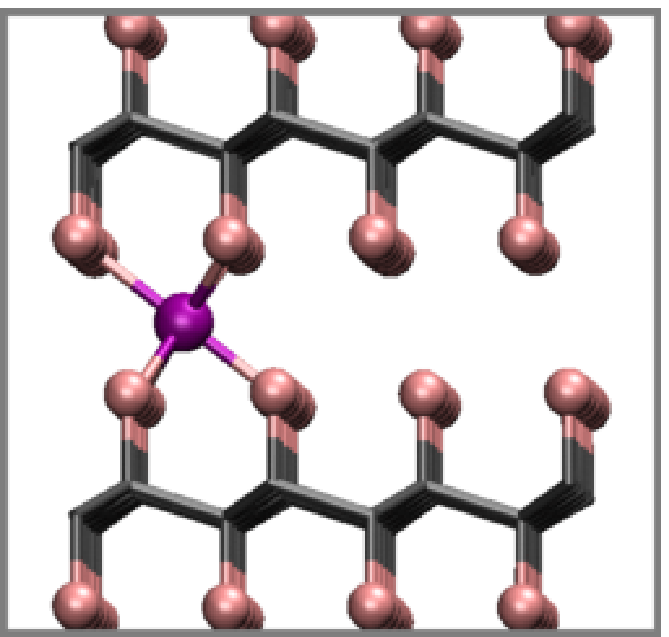} (b)}}
\centerline{\hbox{ \epsfxsize=3.50in \epsfbox{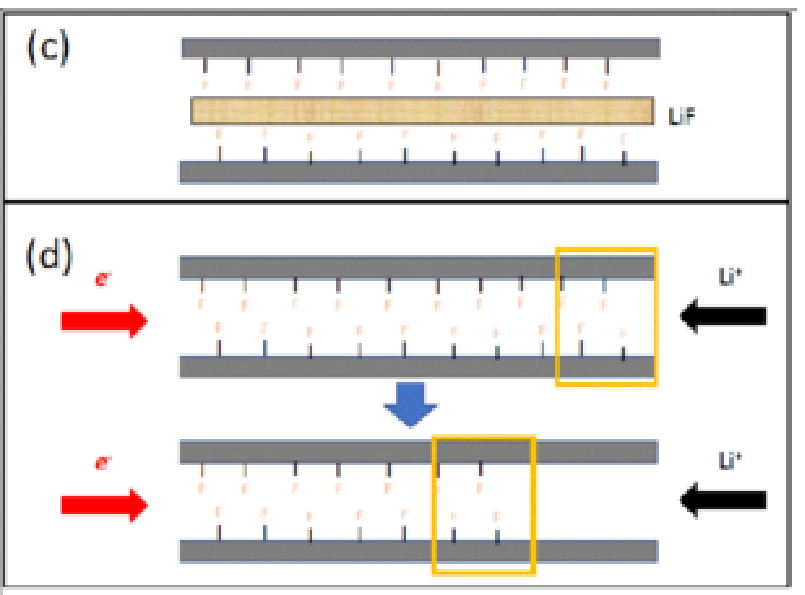} }}
\caption[]
{\label{fig1} \noindent
(a)-(b) Two views of CF$_x$ sheets at $x$=1; zig-zag and arm-chair edges
(in analogy with graphite) are indicated.  A Li-atom is hypothetically
inserted between two CF$_x$ sheets.  
C, F, and Li atoms are depicted as grey, pink, and purple spheres or sticks.
(c)-(d) Schematics of LiF
intermediate phase and edge-propagation CF$_x$ discharge mechanism,
respectively.  Yellow lines highlight local active discharge regions.  
Black sticks are C-F bonds.
}
\end{figure}

Graphite fluoride has the chemical formula CF$_x$ with 0$<$x$\leq$$(1+\delta)$.
When fully fluorinated ($x$$\sim$1), CF$_x$ cathodes exhibit one of the
highest theoertical specific capacities (864 mAh/g) among cathode
materials, \cite{wood1972} and have been commercialized
as primary lithium battery cells.\cite{takeuchi2015,amatucci2007}
CF$_x$ is typically synthesized by enforcing chemical reactions
between F$_2$ gas and graphite or other forms of conductive carbon at
elevated temperatures.\cite{watanabe1980,watanabe1988,hamwi1988,zhong2019}
Idealized models of $x$=1 samples have layers of CF with each carbon atom
chemically bonded to three~other C and one F~atoms (Fig.~\ref{fig1}a-b).  
The registry between adjacent sheets has been predicted to have minimal
effect on the total energy.\cite{goddard2010}  

\begin{figure}
%\centerline{\hbox{ (a) \epsfxsize=3.00in \epsfbox{fig2a.eps} 
%		   (b) \epsfxsize=3.00in \epsfbox{fig2b.ps} }}
%\centerline{\hbox{ (c) \epsfxsize=3.00in \epsfbox{fig2c.ps} 
%		   (d) \epsfxsize=3.00in \epsfbox{fig2d.ps} }}
\centerline{\hbox{ \epsfxsize=4.00in \epsfbox{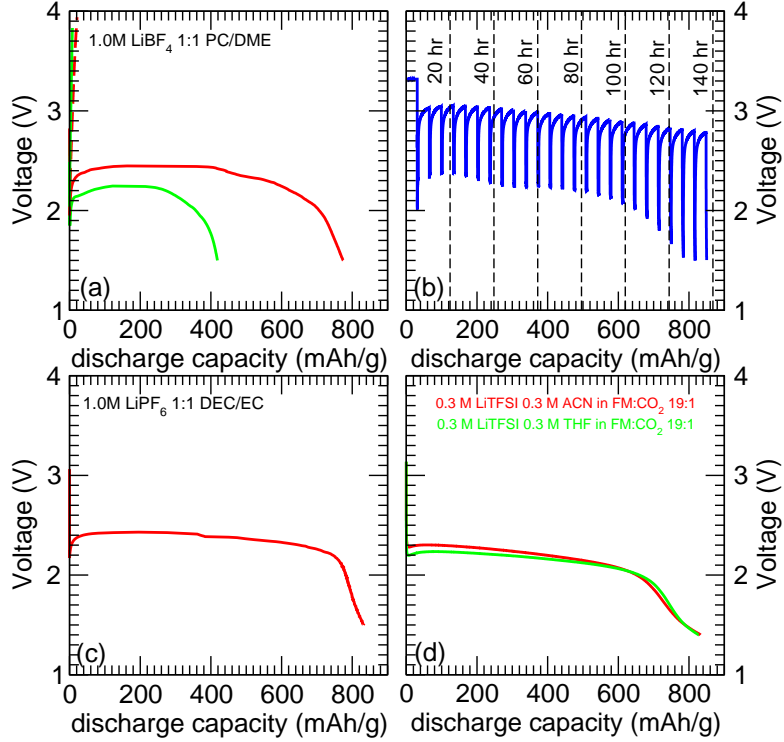} }}
\caption[]
{\label{fig2} \noindent
(a) Galvanostatic discharge and charge of CF cells with PC/DME/Li$^+$/BF$_4^-$ 
electrolyte.  Cells were left at open circuit voltage for 20 hrs then cycled at
a rate of C/20 (red) or C/5 (green). Plot shows that faster rate results
in a lower discharge capacity and that neither cell shows any electrochemical
reversibility (attemped recharge in dashed lines); CF$_{1.09}$ is the initial
composition.  (b) GITT of a cell with the same build as those in
panel (a) CF cell using a 1.789 mA current pulse, corresponding to
a discharge rate of C/20, with a 30 min pulse time and 5 hr rest.  
The final capacity of the GITT measurement is 825.90~mAh/g.  (c) 
Galvanostatic discharge profile with DEC/EC/Li$^+$/PF$_6^-$ electrolyte.
(d) Galvanostatic discharge of CF cells with two liquified gas electrolytes
(LGE) at C/20 rate; CF$_{1.05}$ is the initial composition.
}
\end{figure}

Li/CF$_x$ batteries discharge via the overall reaction 
\begin{equation}
{\rm CF}_x {\rm (s)} + x {\rm Li (s)} \rightarrow {\rm C (s)} + x {\rm LiF (s)} 
	 , \label{eq1}
\end{equation}
where ``(s)'' denotes the solid state.  Using thermodynamic data and
Eq.~\ref{eq1}, the average theoretical voltage is estimated at 4.57~V at $x$=1;
it is even higher at smaller $x$, reaching 5.07~V at $x$=0.7.\cite{wood1972}
In practice,  the usable energy density is significantly lower than
expected from thermodynamics.  Li/CF$_x$ batteries discharge at a plateau
voltage of $<$$\sim$2.5~V at rates of 0.05~C or less
(Fig.~\ref{fig2}a,c,d).  The observed voltage variation is a small fraction of
a volt when using different carbon precursor materials,\cite{carbontypes} or
electrolytes --- including both organic solvents,\cite{solvents,watanabe82}
solid electrolytes,\cite{dudney2014} and liquified gas electrolytes
(Fig.~\ref{fig2}d).\cite{fm0,fm} Galvanostatic intermittent
titration technique (GITT) measurements, which should circumvent most kinetic
limitations, have reported CF$_x$ discharge voltages below 3.1~V
(Fig.~\ref{fig2}b).\cite{gitt}  CF$_x$ materials with more ``ionic'' C-F bonds,
synthesized at $x$ values substantially lower than unity, are reported to
yield slightly higher voltage plateaus and higher discharge rates at the expense
of lower overall capacities.\cite{lirecharge2,lirecharge3,yoshida}
Disordered/nanoscale carbon precurors also yield rate capability
improvement.\cite{ultrafast,lirecharge1}  Unlike Li/CF$_x$,
Na/CF$_x$ batteries have been demonstrated to be
rechargeable.\cite{narecharge0,narecharge1,narecharge2,shao2016}
%Some of the charging profiles exhibit typical battery-like
%signatures.\cite{narecharge0,narecharge1,shao2016}

Achieving performance near the ideal theoretical values as predicted by
thermodynamics will significantly broaden the application space and impact
of CF$_x$ batteries. A detailed elucidation of the CF$_x$
discharge mechanism at the atomic length-scale is required in order to
ultimately achieve increased voltage windows, higher energy
densities, higher power capability, and improved rechargeability.
Several discharge mechanisms have been proposed to explain the voltage
profiles; they differ by whether intermediate phases, edge planes, and/or
solvent molecules are involved.  

{\it Mechanism A: Two Phase Behavior}
In the absence of intermediate phases, the discharge should exhibit
two-phase behavior, and the observed voltage should follow
Eq.~\ref{eq1}.\cite{coreshell,read2011}  The reason for the discrepancy
between the thermodynamic ($>$4.57~V) and observed ($\sim$2.5~V)
discharge voltages is then ascribed to slow kinetics and/or nanosize
effects associated with LiF/NaF products.  One argument against such a
two-phase behavior is that Eq.~\ref{eq1} predicts an increase
of the equilibrium voltage ${\cal V}_i$ as discharge proceeds ($x$
decreases),\cite{wood1972} which has not been observed (Fig.~\ref{fig2}).

{\it Mechanism B: Intermediate Phase} Alternatively,
an intermediate phase has been invoked to explain the 2.5~V plateau.
This may be a ternary phase such as CLi$_x$F,\cite{whittingham} which can be
a thin sheet of one or more LiF layers intercalated between CF$_x$ or C sheets
(Fig.~\ref{fig1}c).\cite{whittingham,shao2016} In this case, Eq.~\ref{eq1}
does not govern the discharge voltage.  Changing CF$_x$ lattice
constants with discharge have been reported in {\it in situ} X-ray
analysis,\cite{manny2007,manny2007a} which may suggest such a LiF
intercalation structure.  However, Mechanism (B) appears less consistent with
CF$_x$ not in stack-like configurations, e.g., those synthesized using 
carbon nanotube precursors.  

{\it Mechanism C: Edge-Mediated} Another candidate for the intermediate
phase is a solvent-coordinated Li$^+$ complex,\cite{amatucci2007,nakajima1999}
\begin{equation} 
{\rm Li} + {\rm CF} + x{\rm S} \rightarrow 
	({\rm CF}^-)( {\rm Li}^+) ({\rm S})_x , \label{eq2}
\end{equation}
where ``S'' is a solvent molecule.  Such a mechanism necessarily requires an
edge-propagation rather than bulk-phase reaction pathway.  Edge-propagation
mechanisms are attractive because they would be consistent with an extended
plateau voltage region independent of the extent of the state-of-charge ($x$).
This is because if the discharge behavior only depends on the local
configuration (Fig.~\ref{fig1}d), to a first approximation the spatial location
of discharge, dependent on $x$, would not affect the discharge voltage.
This is consistent with the appearance of a near constant voltage
discharge plateau (Fig.~\ref{fig2}a).
However, the intermediate phase associated with this mechanism would be small
and hard to detect via X-ray diffraction.\cite{manny2007,manny2007a}

Mechanisms (B) and (C) are not mutually exclusive.  Strictly speaking, 
Mechanism (B) does not completely specify a discharge pathway, in terms
of the order in which C-F bonds are broken.  If thermodynamic equilibrium is
assumed, i.e., C-F bonds are broken in ascending order of bond energies, 
Mechanism (B) may be consistent with an overall, lower average discharge
voltage than 4.57~V, but would fail to explain why the observed discharge
voltage does not increase with decreasing $x$ (Fig.~\ref{fig2}), like
thermodynamics would predict.\cite{wood1972} However, when combined with
Mechanism (C), the formation of a CLi$_x$F phase (Fig.~\ref{fig1}c) may follow
edge-mediated C-F bond-breaking events.

So far, purely experimental efforts have not definitively determined
the mechanism.  Computational work will shed light on Mechanisms (A)-(C).
Regarding Mechanism (A), two-phase-like solid state conversion reactions are
well understood and routinely modeled using a phase-diagram
approach.\cite{conversion1,conversion2}  The effect of nanosized
charge/discharge products has been addressed within the framework of phase
diagram calculations to explain the discrepancy between bulk phase
thermodynamics predictions and observed battery discharge
profiles.\cite{shao2016,mengnano}  However, nanosizing is usually insignificant
for particle size on the order of 10~nm, which have been reported as the
dimensions of LiF discharge products,\cite{read2011} although further
experimental work is needed to ascertain the size distribution.
Furthermore, the phase diagram approach
assumes equilibrium conditions and reversible reactions.  While this is true
of many conversion cathode materials,\cite{conversion1,conversion2}
rechargeability associated with Eq.~\ref{eq1} has yet to be demonstrated
in any electrolyte, which suggests that equilibrium conditions do not apply.
In the SI (Sec.~S2), we further report that Density Functional Theory (DFT)
electronic structure calculations predict a barrier for the exchange of two
neighboring F-vacancies on a CF sheet that is far
too large to permit diffusion, consistent with previous bond-strength
calculations.\cite{dubecky2015,venkat2020} This shows that different C-F bond
configurations in CF$_x$ materials are not at equilibration with each other.

Regarding Mechanisms (B) and (C): in this work, we apply DFT to show that
Mechanism (C),
even without solvent molecules, gives good agreement with experiments.
Motivated by edge-initiated discharge in graphite used as lithium ion battery
anodes,\cite{tateyama2020,vedge,borodin2012} 
we examine edge-propagation CF$_x$ discharge (Fig.~\ref{fig1}d).  We find
that Li or Na intercalation into interfacial sites between insulating CF and
conductive, defluorinated graphite regions constitutes a small  ``intermediate
phase'' most consistent with the observed CF$_x$ discharge voltage plateau
(Fig.~\ref{fig2}).  This intermediate dovetails with
classic ``interfacial charge storage'' behavior.\cite{maier2014,maier2007}   
Li insertion is followed by defluorination at the CF/C interface and then
further Li insertion, leading to quasi-one-dimensional, row-by-row
defluorination and subsequent formation of LiF.  LiF formation energetics
does not determine the voltage.  In addition to CF$_x$ stacks, our model is
applicable to CF$_x$ flakes and fluorinated carbon nanotubes with small
curvatures, as long as their discharge involves row-by-row defluorination.  
In this work, we adopt idealized, defect free, partially defluorinated
CF$_x$ with zig-zag and arm-chair edges.  Defects in carbon sheets and other
heterogeneity (e.g., C-F vs.~C-F$_2$ distributions\cite{nmr1,nmr2,nmr3})
also affect CF$_x$ battery operations; these complexities will be deferred to
future modeling work.

Our calculations also consider Li/CF$_x$ recharge and discharge rates.
Regarding solvent effects: given the large variation in the binding energies
between Li$^+$ and different solvent molecules we will discuss, the solvent
dependence of discharge voltage reported in the literature, on the order
of 0.2~V, appears surprisingly small and requires further
elucidation.\cite{nakajima1999,watanabe1982}  This will be addressed by
adding solvent molecules at CF$_x$ edge sites in DFT simulations, and
comparing to our measurements done in two solvents.  A significant amount
of DFT modeling work on CF$_x$ have been
reported.\cite{boukhalov2016,charlier1993,goddard2010,bettinger2004,sahin2011,takagi2002,karlicky2012,zhou2014,yakobson2012,fan2020,ouyang2015,venkat2020}
Few of them deal with CF$_x$ edges; however, this prior research provides
significant guidance and starting structures for the work discussed below.  

%The CF$_x$ cathode open circuit voltage (OCV) has been reported at 
%$\sim$3.4~V.\cite{manny2007}  

\section{Method}

\begin{table}\centering
\begin{tabular}{c|r|r|l|r} \hline
system & dimensions & stoichiometry & $k$-sampling & Figure \\ \hline
zig-zag &  6.10$\times$2.60$\times$32.00 & C$_{18}$F$_{20}$ &
          3$\times$10$\times$1  & \\
zig-zag &  6.10$\times$5.21$\times$32.00 & C$_{36}$F$_{40}$ &
          3$\times$5$\times$1  &   \\
zig-zag &  6.10$\times$7.81$\times$32.00 & C$_{54}$F$_{60}$ &
          3$\times$3$\times$1  &   \\
arm-chair &  6.10$\times$4.63$\times$32.00 & C$_{36}$F$_{40}$ &
          3$\times$6$\times$1  & \\
arm-chair &  6.10$\times$9.65$\times$32.00 & $_{72}$F$_{80}$ &
          3$\times$3$\times$1  &   \\
zig-zag/Li$^+$ &  12.20$\times$10.40$\times$32.00 & C$_{144}$F$_{128}$Li &
          2$\times$5$\times$1  & Fig.~\ref{fig3}a \\
zig-zag/Li$^+$ &  12.20$\times$10.40$\times$32.00 & C$_{144}$F$_{80}$Li &
          2$\times$5$\times$1  & Fig.~\ref{fig3}f \\
zig-zag/Li$^+$ &  12.20$\times$20.80$\times$32.00 & C$_{288}$F$_{256}$Li &
          2$\times$5$\times$1  &   \\
arm-chair/Li$^+$ &  12.20$\times$9.27$\times$36.00 & C$_{144}$F$_{112}$Li &
          2$\times$3$\times$1  & Fig.~\ref{fig6}a \\
arm-chair/Li$^+$ &  12.20$\times$9.27$\times$44.00 & 
	  Au$_{\rm 40}$C$_{144}$F$_{112}$Li &
          2$\times$3$\times$1  & Fig.~\ref{fig6}b \\
zig-zag/Li$^+$ & 12.20$\times$10.40$\times$44.00 & C$_{144}$F$_{200}$Li$_{72}$ &
          2$\times$5$\times$1  & Fig.~\ref{fig7}a \\
\hline
\end{tabular}
\caption[]
{\label{table2} \noindent
Computational details of representative simulation cells.  The dimensions
are in units of \AA$^3$.
}
\end{table}

To calculate energies associated with voltages (Eq.~\ref{eq1}-\ref{eq2}),
we apply static DFT calculations with periodically
replicated simulation cells, the Vienna Atomic Simulation Package (VASP)
version 5.3,\cite{vasp1,vasp1a,vasp2,vasp3} and the Perdew-Burke-Ernzerhof
(PBE) functional.\cite{pbe}  
A 400~eV planewave energy cutoff is imposed, except that a 500~eV cutoff
is used when optimizing simulation cell sizes.  We adopt a 6.1~\AA\,
CF$_x$ inter-sheet spacing, similar to Ref.~\onlinecite{goddard2010}.

Our DFT simulation cells represent stacked, partially defluorinated
CF$_x$ sheets with sharp interfaces between the fully defluorinated
graphene and fully fluorinated (CF) regions.  The simulation cell sizes
correspond to CF, not graphite, lattice constants; hence strain
develops in the defluorinated region.\cite{boukhalov2016} 
Representative simulation cell dimensions, stoichiometries, and
Brillouin zone sampling settings are listed in Table~\ref{table2}.  Other
calculations involve variations on these cells.  All simulation cells
considered are overall charge-neutral.  The dipole moment correction is
applied in all but a few calculations which do not contain a vacuum
region.\cite{dipole}  This correction only avoids charge/image-charge
interactions in the $z$ direction perpendicular to the CF/C interface.
When the simulation cell exhibits significant dipole moments, systematic
increase of cell size in the lateral directions is still necessary to
converge relevant energy differences.
Spin-polarized DFT is used except for certain arm-chair edge calculations
with even number of electrons, where non-spin-polarized DFT gives the same
result.  A few calculations apply the generally more accurate DFT/HSE06
functional.\cite{hse06a,hse06b,hse06c}  The dispersion-corrected optB86b-vdW
functional is also tested in some cases.\cite{dftvdw}  The electronic voltage
is determined using the work function approach, which is possible in the
absence of liquid electrolytes.\cite{pccp}

We also perform finite temperature
{\it ab initio} molecular dynamics (AIMD) simulations of a small Li cluster
in contact with zig-zag or arm-chair edges.  These are short-circuit
condition simulations which help motivate the sharp CF/C boundaries used in
T=0~K DFT simulation cells.  A Nose thermostat imposes T=350~K conditions.
The simulations adopt $\Gamma$-point Brillouin zone sampling; other settings
are the same as those discussed above.  The zig-zag edge AIMD cell has a
17.93$\times$13.01$\times$28~\AA$^3$ dimension and a
C$_{150}$F$_{180}$Li$_{21}$ stoichiomtry.  The arm-chair edge simulation cell
has a 18.30$\times$13.51$\times$28~\AA$^3$ dimension and a
C$_{126}$F$_{144}$Li$_{21}$ stoichiometry.  The results are described in the SI.

Experimental details are discussed in the SI.

\section{Results}

\subsection{Experimental Results}
\label{gitt} 
Discharge and GITT profiles for CF$_x$ cells with propylene carbonate
(PC)/1,2-dimethoxyethane
(DME)/lithium tetrafluoroborate (LiBF$_4$) electrolytes are
depicted in Fig.~\ref{fig2}a-b.  They are similar to results reported in the
literature for Li/CF$_x$ batteries.\cite{gitt1,gitt2} Fig.~\ref{fig2}a shows
that the discharge voltage profile exhibits a plateau around 2.4~V at the
C/20 rate; discharge voltage and capacity become much more limited at the
faster C/5 rate.   Recharging does not occur at either rate when an upper
voltage cut-off of 4.0~V is imposed.  The GITT results (Fig.~\ref{fig2}b)
show that the highest observed discharge voltage after the initial pulse
is $\sim$3.05~V.  The instantaneous diffusion constants associated with
Fig.~\ref{fig2}b are depicted in the SI~(Fig.~S1).  Omitting the
first two pulses, we estimate that the average Li$^+$ diffusion constant
to be 4.47$\times$10$^{-12}$~cm$^2$/s assuming Li$^+$ enters the cathode
via an ion-insertion pathway (not just a surface reaction).  This value
is two to three orders of magnitude slower than in commercial secondary
lithium-ion battery materials, but does not appear forbiddingly
slow.\cite{diffus1}  
The galvanostatic discharge profile of an electrolyte with ethylene carbonate
(EC)/diethyl carbonate (DEC)/lithium hexafluorophosphate (LiPF$_6$)
(Fig.~\ref{fig2}c) is similar to that without EC (Fig.~\ref{fig2}a).

Galvanostatic discharge data of two fluoromethane (FM)-based liquified
gas electrolytes (LGE), with lithium bis(trisfluoromethanesulfonyl)imide
(LiTFSI) salt, CO$_2$ additive, and either
0.3~M acetonitrile (ACN) or 0.3~M tetrahydrofuran (THF), show lower
discharge voltages compared with carbonate-based liquid electrolyte at
the same C/20 rate (Fig.~\ref{fig2}d).  This solvent variation will be
compared with DFT studies below.

\subsection{Intrinsic Defluorination Thermodynamics Does not Explain Voltage}

The equilibrium voltage ${\cal V}_i$ associated with a small defluorination
increment $\delta x$ along a particular defluroination pathway,
\begin{equation}
{\rm C}{\rm F}_{x}{\rm (s)} + \delta x {\rm Li(s)} \rightarrow 
{\rm CF}_{x-\delta x} {\rm (s)} + \delta x {\rm LiF(s)} \hspace*{0.1in} 
		, \label{eq3}
\end{equation}
is given by
\begin{equation}
|e| {\cal V}_i =  -[ E({\rm C}{\rm F}_{x}) - E({\rm CF}_{x-\delta x})]/\delta x
		+ E({\rm LiF}) - E({\rm Li}) ,\label{eq4}
\end{equation}
where the energies $E()$ all refer to those of solid phases. The larger a C-F
bond energy $([ E({\rm C}{\rm F}_{x}) - E({\rm CF}_{x-\delta x})]/\delta x)$
for a particular carbon atom, the lower is the voltage required to defluorinate
that atom.  Note that more positive ${\cal V}_i$ and more negative energy
changes mean more favorable reactions with Li.  We stress that ${\cal V}_i$ can
only be realized in experiments if the corresponding reaction step occurs
under equilibrium conditions.  

\begin{figure}
\centerline{\hbox{ (a) \epsfxsize=1.50in \epsfbox{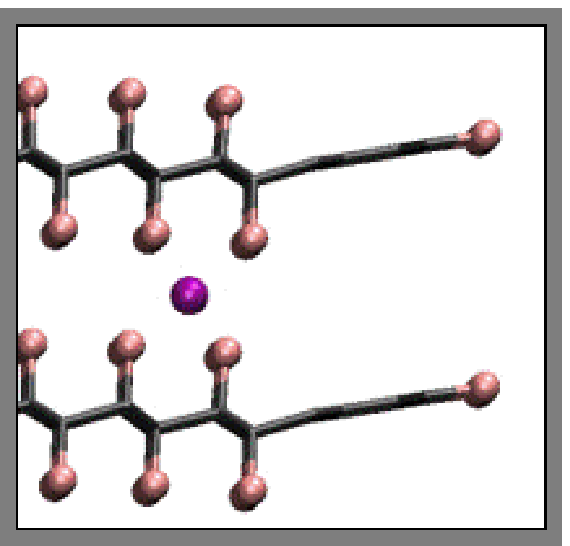} 
		   (c) \epsfxsize=1.50in \epsfbox{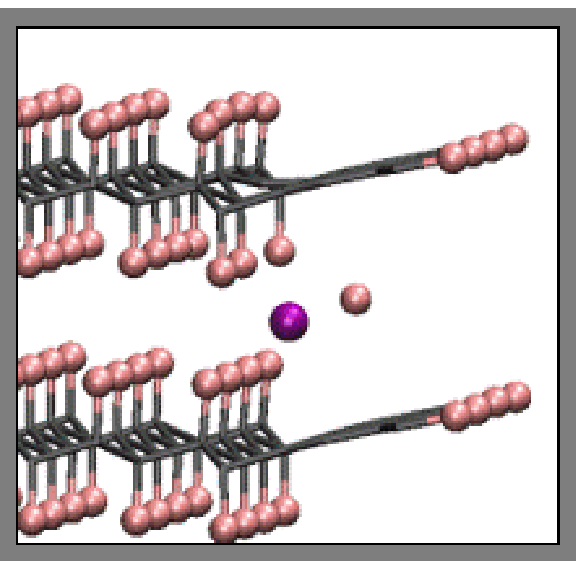} 
		   (e) \epsfxsize=1.50in \epsfbox{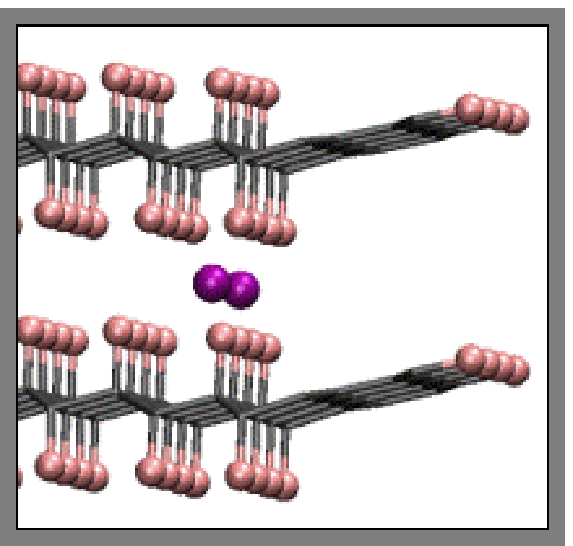} }}
\centerline{\hbox{ (b) \epsfxsize=1.50in \epsfbox{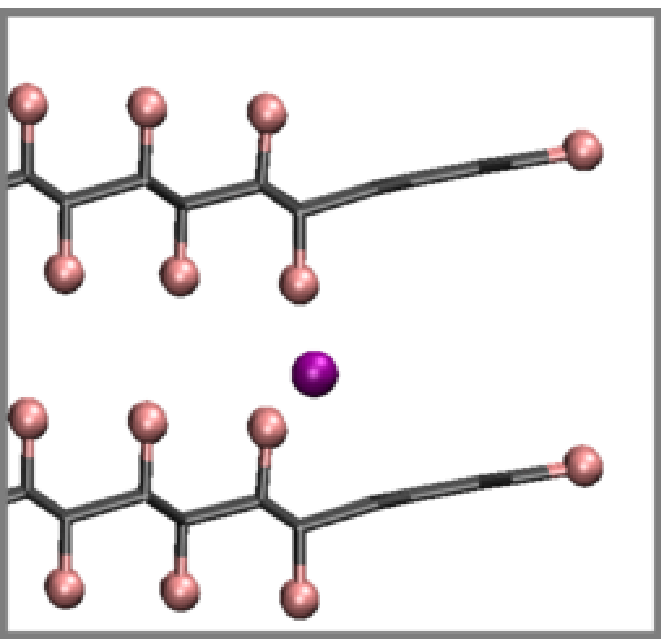} 
		   (d) \epsfxsize=1.50in \epsfbox{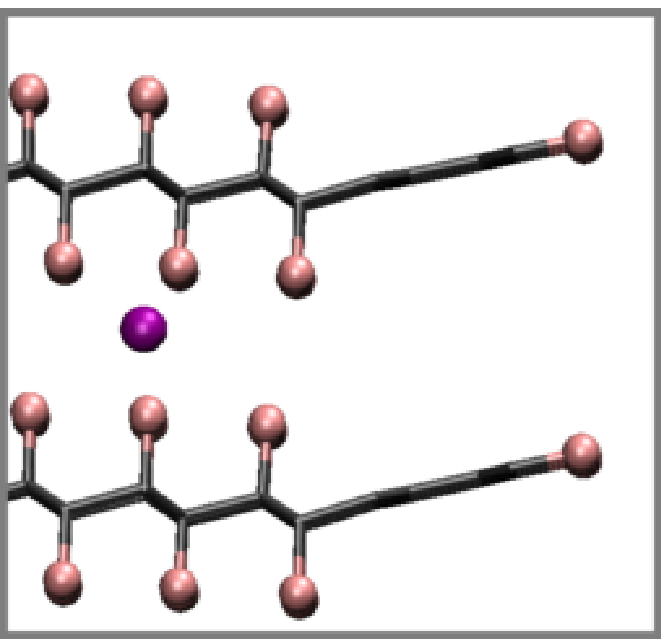} 
		   (f) \epsfxsize=1.50in \epsfbox{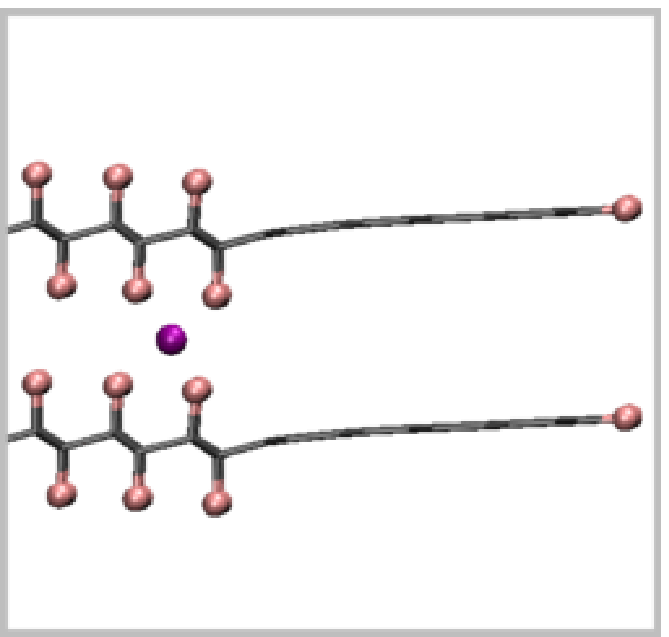} }}
\caption[]
{\label{fig3} \noindent
(a) 6-coordinated CF/C Li ``interfacial site.''
(b) 3-coordinated Li ``outer binding site.'' 
(c) Breaking a C-F bond in panel (b).
(d) 6-coordinate Li-site inside the CF/C interface.
(e) Inserting two Li at nearby CF/C interfacial sites.
(f) CF/C interface site after further removal of several rows of F atoms from
panel (a).  For color key, see Fig.~\ref{fig1}.
}
\end{figure}

First we discuss the energetics of bulk phase CF$_x$ defluorination without
metal content in the simulation cell.  In this formulation, the effect of Li$^+$
is solely manifested in the LiF formation energy (Eq.~\ref{eq4}).
Similar calculations have been applied in previous DFT studies of CF$_x$ batteries.\cite{boukhalov2016,charlier1993,goddard2010,bettinger2004,sahin2011,takagi2002,karlicky2012,zhou2014,yakobson2012,ouyang2015}
No CF$_x$ edge or solvent molecule is included in this approach.
Using the PBE, HSE06, and optB86b-vdW functionals, the equilibrium voltages
associated with Eq.~\ref{eq1} at $x$=1 (i.e., averaging the voltage
defluorinating the entire CF material) are found to be ${\cal V}_i$=4.38, 4.40,
and 5.16 V, respectively.  Periodically replicated CF$_x$ or graphene sheets
instead of 3-dimensional structures for both ``CF'' and ``C'' in Eq.~\ref{eq1}
yield similar results, except that the optB86b-vdW functional voltage is
slightly lowered to 5.11~V.  The PBE and HSE06 functionals are in closest
agreement with thermodynamic data,\cite{wood1972} and the computationally
less costly PBE is applied for the remainder of this work.

We also consider starting at the $x$$\rightarrow$1 limit and removing two
neighboring F atoms from a periodically replicated CF sheet via Eq.~\ref{eq5}.
A ${\rm C}_{48}{\rm F}_{48}$ model is used (i.e., $\delta x$=1/24,
defluorinating a small portion of the CF).
Eq.~\ref{eq3} coupled with Eq.~\ref{eq4} yields ${\cal V}_i$=2.86~V, which
is significantly lower than the average value of 4.38~V.  See
Table~\ref{table1}.  This PBE-predicted
trend, whereby ${\cal V}_i$ increases with state of discharge (smaller $x$), is
qualitatively consistent with thermodynamic data,\cite{wood1972} but is
inconsistent with observed electrochemical discharge voltage profiles
(Fig.~\ref{fig2}).  We have not considered intermediate values of $x$
partly because of the multiplicity of possible defluorination
configurations with complex energy landscapes,\cite{boukhalov2016} and
partly because our edge-propagation mechanism described below provides a
more physical, quasi-kinetic pathway for the sequence of F-atom to be
removed, than a global optimal energy criterion.

\begin{table}\centering
\begin{tabular}{ l l l l } \hline
system & ${\cal V}_i$ & figure & notes \\ \hline
bulk CF$_x$ & 4.38~V & NA & averaged over $x$$<$1 \\
bulk CF$_x$ & 2.86~V & NA & at $x$=1 only \\
bulk CF$_x$+Li & 0.93~V  & Fig.~\ref{fig1}a-b & insert Li \\
zig-zag & 2.62~V  & Fig.~\ref{fig3}a & insert Li \\
zig-zag & 1.45~V  & Fig.~\ref{fig3}e & insert 2nd Li \\
zig-zag & 2.56~V  & Fig.~\ref{fig3}f & further defluorination \\
zig-zag & 2.90~V  & Fig.~\ref{fig5}a & with EC \\
zig-zag & 2.49~V  & Fig.~\ref{fig5}b & with FM \\
zig-zag & 2.66~V  & Fig.~\ref{fig5}d & sheet, with EC \\
zig-zag & 2.25~V  & Fig.~\ref{fig5}f & sheet, with FM \\
arm-chair & 1.49~V  & Fig.~\ref{fig6}a & insert Li \\
arm-chair & 1.81~V  & Fig.~\ref{fig6}b & with Au(111) \\ \hline
\end{tabular}
\caption[]
{\label{table1} \noindent
Computed voltages (${\cal V}_i$) at different configurations.
Bulk defluorination voltages follow Eq.~\ref{eq1} or~\ref{eq4}
while Li-insertion voltages follow Eq.~\ref{eq5}. EC and FM are
ethylene carbonate and fluoromethane, respectively.
}
\end{table}

\subsection{Li-insertion Energetics at Zig-Zag CF/C Interface}
\label{edge1}

During CF$_x$ discharge, Li$^+$ should be present at the CF$_x$ edge facing
the electrolyte (right side of Fig.~\ref{fig1}d).  Simultaneously, an $e^-$ is
injected from the current collector (not explicitly included in our DFT model)
on the left side.  The added $e^-$ moves infinitely fast in ground state
DFT calculations; therefore operationally these two charge injection steps
are consistent with adding a Li atom on the surface facing the electrolyte.  In
the SI (Sec.~S7), we describe {\it ab initio} molecular dynamics simulations of
Li nanoclusters in contact with and reacting with CF$_x$ edges under
short-circuit conditions.  These are qualitative in nature, aimed to illustrate
what can spontaneously occur under artificially accelerated conditions. 

In this section we consider inserting one Li at a time, which is more relevant
to slow discharge rate conditions seen in practical CF$_x$ cells.
Fig.~\ref{fig3}a depicts such a Li at the zig-zag
edge interface between CF and the defluorinated graphite region of partially
defluorinated CF$_x$.   The existence of sharp
CF/C boundaries is motivated by our hypothesis that an edge-propagation
mechanism would not significantly depend F Li-content ($x$ in CF$_x$,
Fig.~\ref{fig1}d), and to a lesser extent by AIMD simulation results
(SI).  \color{black}  At this ``CF/C interfacial site,'' Li
is coordinated to six C-F groups.  

Unlike the formation of LiF discharge products, which involves typically
slower nucleation events, Li$^+$ insertion into CF$_x$
is diffusive, assisted by electric fields.  It should be fast
(Sec.~\ref{gitt})\cite{nakajima1999} and reversible.  This equilibrium
assumption allows the use of
\begin{equation}
|e| {\cal V}_i =  [ E({\rm C}{\rm F}{\rm Li}_x) - E({\rm C}{\rm F}
	{\rm Li}_{x-\delta x})]/\delta x  - E({\rm Li}) ,\label{eq5}
\end{equation}
to describe the observed voltage.  Eq.~\ref{eq5} would be identical to
expressions used to calculate LiC$_6$ equilibrium voltages if ``CF'' is 
replaced with ``C.''  In this formulation, LiF(s) formation subsequent to Li
insertion does not figure into ${\cal V}_i$, unlike in most previous DFT work.
We find that Li insertion at the interfacial site (Fig.~\ref{fig3}a) yields
${\cal V}_i$=2.62~V via Eq.~\ref{eq4}.  This value is in good agreement with
the observed CF$_x$ discharge plateau value of $\sim$2.5~V (Fig.~\ref{fig1}a).
GITT measurements, which should give voltages closer to equilibrium values,
yield open circuit voltage values higher by only 0.4-0.5~V.  This covers
the ${\cal V}_i$ range predicted.

\begin{figure}
\centerline{\hbox{ \epsfxsize=4.00in \epsfbox{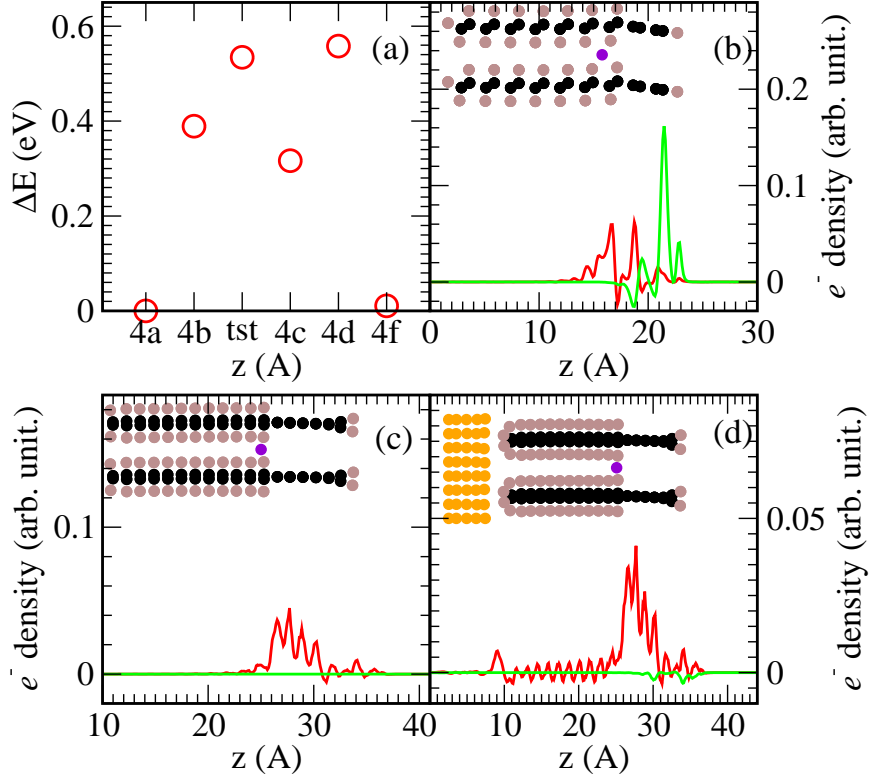} }}
\caption[]
{\label{fig4} \noindent
(a) Energy landscape for Fig.~\ref{fig3} [recall panel (f) refers to
a different F-content]. (b)-(d)
Differential charge (red) and spin (green) densities before/after removing
a Li atom.  (b) Zig-zag edge; the Li is at $z$=15.8~\AA.  (c) Arm-chair CF/C
interfaces; the Li is at $z$=15.2~\AA.  (d) Same as (c) but with an Au(111)
slab on the left side; Li is at $z$=25.2~\AA.  For color key,
see Fig.~\ref{fig1}.
}
\end{figure}

This value of 2.62~V is surprising in light of the less favorable Li
insertion into bulk CF, both fluorinated and defluorinated.  Li$^+$
intercalation into graphite (as model for defluorinated CF$_x$) has been
extensively studied in the context of graphite anodes.  The process is fast and
reversible, even though an electron-insulating but ion-conducting solid
electrolyte interphase film is present.\cite{abe2004,borodin2012}  
This reaction,
\begin{equation}
{\rm Li(s)} + 6 {\rm C (s)} \rightarrow {\rm LiC}_6 {\rm (s)} , \label{eq6}
\end{equation}
occurs at ${\cal V}_i$=0.1-0.2~V vs.~Li$^+$/Li(s), which is far lower than
the observed CF$_x$ discharge voltages (Fig.~\ref{fig2}).  In contrast,
the equilibrium voltages associated with Li-insertion into CF$_x$ stacks
(Fig.~\ref{fig1}a),
\begin{equation}
 {\rm Li(s)} + (1/x) {\rm CF (s)} \rightarrow (1/x){\rm Li}_x{\rm CF}{\rm (s)},
\label{eq7}
\end{equation}
have not been reported.  Our DFT/PBE calculations in the dilute limit
($x$=1/48) yield 0.93~V vs.~Li$^+$/Li(s).  This ${\cal V}_i$ is higher than
that associated with LiC$_6$, but is significantly lower than experimental
discharge voltages (Fig.~\ref{fig2}).  Since the excess $e^-$ from Li
insertion will reside in the conduction band of the insulating CF, the
0.93~V value is likely overestimated because the PBE functional underestimates
band gaps.  This value is consistent with the defect-free single CF sheet
voltage of $\sim$1.0~V reported in Ref.~\onlinecite{fan2020} after adjusting
the Li atom binding energy used therein with the lithium metal cohesive energy
reference.

The anomalously high ${\cal V}_i$ at the interface compared to bulk phases
is a manifestation of interfacial charge storage
behavior.\cite{maier2014,maier2007}  Li$^+$ favors insertion into the CF region,
where it is stabilized by CF polar groups, while the accompanying $e^-$ is
partially delocalized in the nearby metallic carbon region.  This charge
sharing principle is illustrated in the differential charge and spin density
plot (Fig.~\ref{fig4}b), computed by removing the inserted Li atom 
while freezing all other atoms.
In contrast, while graphite is metallic and readily accommodates an excess
$e^-$, it interacts weakly with Li$^+$, resulting in a low overall
Li-binding energy and a low ${\cal V}_i$.  For CF, the interaction between
Li$^+$ and the polar C-F bond is more energetically favorable than that
between Li$^+$ and graphite.  However, CF$_x$ is an insulator at large $x$,
and adding $e^-$ to its conduction band is unfavorable.  Hence the interface
between CF and defluorinated CF regions is uniquely suited to inserting Li.

\subsection{Subsequent C-F Bond-Breaking Kinetics}

\begin{figure}
\centerline{\hbox{ (a) \epsfxsize=2.50in \epsfbox{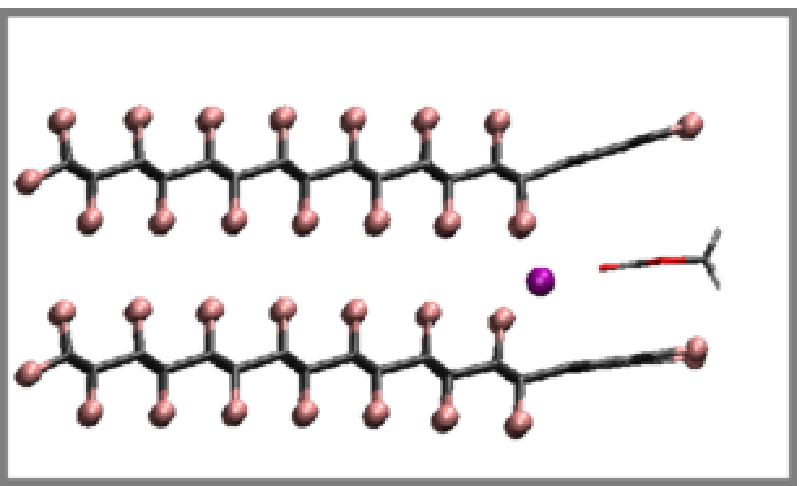} 
		   \epsfxsize=2.50in \epsfbox{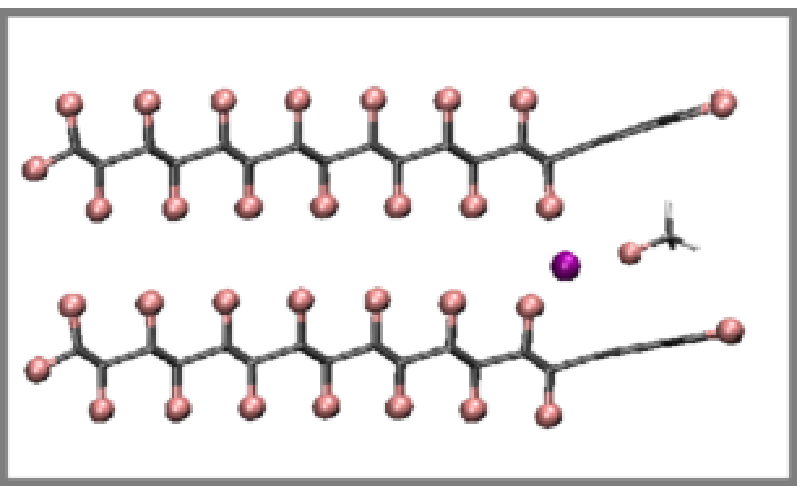} (b)}}
\centerline{\hbox{ (c) \epsfxsize=2.50in \epsfbox{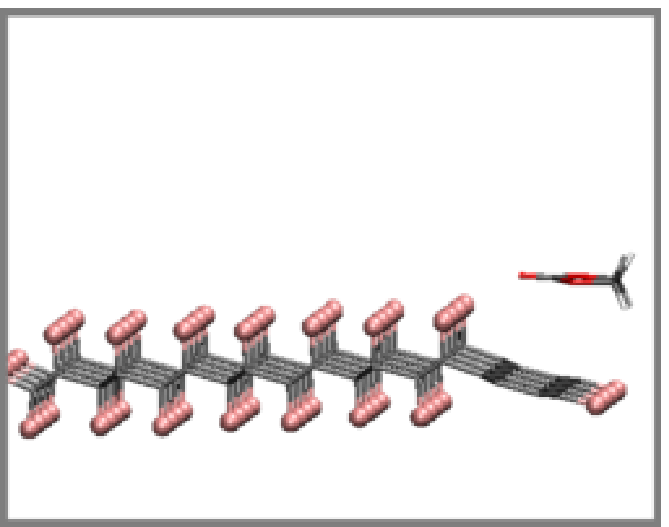} 
		   \epsfxsize=2.50in \epsfbox{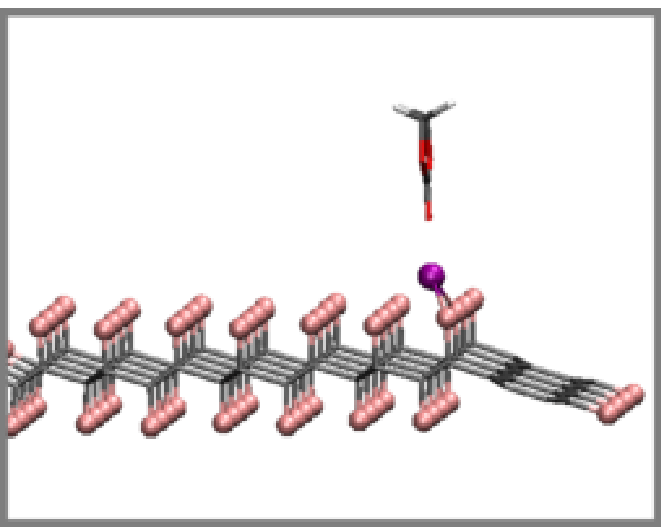} (d)}}
\centerline{\hbox{ (e) \epsfxsize=2.50in \epsfbox{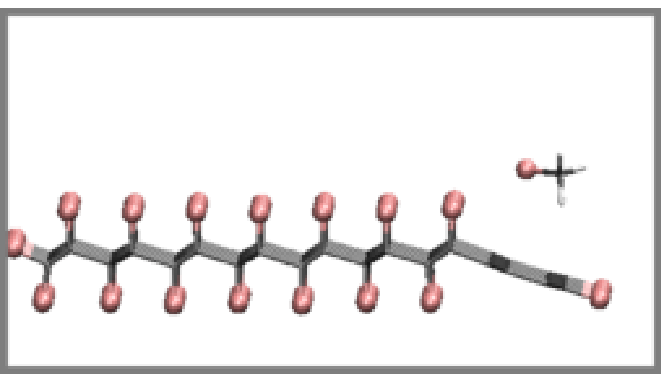} 
		   \epsfxsize=2.50in \epsfbox{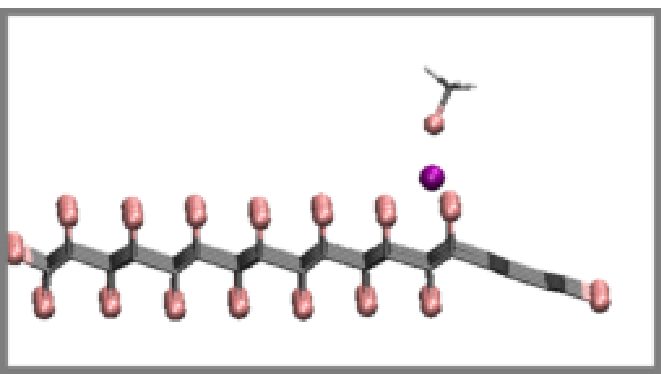} (f)}}
\caption[]
{\label{fig5} \noindent
(a)-(b): Adding a single EC or FM solvent molecule to Li$^+$ at the
3-coordinated outer binding site at the zig-zag edge.
(c): EC on isolated partially defluorinated CF$_x$ sheet.  
(d): Same as (c) but with Li$^+$ coordinated to 3 CF bonds and the EC molecule.
(e): FM on isolated partially defluorinated CF$_x$ sheet.  
(f): Same as (e) but with Li$^+$ coordinated to 3 CF bonds and the FM molecule.
${\cal V}_i$ associated with panels (a), (b), (d), and (f) are 2.90~V, 2.49 V,
2.66~V, and 2.25~V, respectively.  The color key is as in Fig.~\ref{fig1};
in addition red and white sticks represent O and H atoms.
}
\end{figure}

Next we consider the steps subsequent to Li insertion.
We move the inserted Li to a 3-coordinated surface site outside
the interfacial site and re-optimize the configuration (Fig.~\ref{fig3}b).
This will be referred to as  the ``outer binding site.''  The energy
associated with this site is a modest 0.39~eV higher than the interfacial
site (Fig.~\ref{fig3}a).  The subsequent C-F bond breaking event 
(Fig.~\ref{fig3}c), a prerequisite to LiF crystal formation, is exothermic
by 0.07~eV relative to Fig.~\ref{fig3}b.  It is endothermic relative
to Fig.~\ref{fig3}a by 0.32~eV, but that enthalpy cost is compensated by
the canonical $\sim$0.4~eV entropy gained by releasing a ``molecule'' (LiF) at
T=300~K, \color{black} due to the transformation of low-entropy vibrational
degrees of freedom in a configuration with an intact C-F bond to the
high-entropy rotational and translational degrees of freedom after the
bond is broken.\cite{copco2} The value of $\sim$0.4~eV assumes that the
system is at equilibrium and the product is at a 1.0~M concentration
in the liquid phase.  In reality, the LiF product is continuously consumed
and removed from the electrolyte, so 0.4~eV is a lower-bound rough 
estimate (SI Sec.~S6).  
The activation energy ($\Delta E^*$) associated with this
defluorination step is a modest 0.53~eV relative to Fig.~\ref{fig3}a.
Assuming a standard kinetic prefactor of 10$^{12}$/s, once Li is inserted,
F$^-$ will be released in millisecond time scales at T=300~K.

Over time, the released LiF diatomic fragment is expected to nucleate
with other LiF units to form LiF nanocrystalline discharge products.
Experimentally, crystallites of sizes as small as 30~nm have been inferred
These crystallites have been observed to be $>$10~nm in size,\cite{read2011}
bulk like for the purpose of calculating energetics, though more rigorous
experimentation is needed to conclusively determine the final LiF dimensions
in discharged CFx cathodes.  Our DFT/PBE
calculations show that LiF(g)$\rightarrow$LiF(s), where ``(g)'' and ``(s)''
stands for gas and crystalline solid phases respectively, is exothermic by
2.66~eV.  Therefore LiF nanocrystal formation is not reversible unless
$>$2.66~eV external energy is injected, for example, via applying a high
potential (Sec.~\ref{recharge}).  This may help explain why Li/CF$_x$ cells
have so far been non-rechargeable.  \color{black} We have not considered the
kinetics or possible overpotentials associated with the subsequent nucleation
or growth of LiF crystals from these 2-atom LiF units.  However, a reactive
molecular dynamics (MD) approach has previously shown that such growth can
occur in MD time scales ($\ll$1~second), in a different battery
system.\cite{garofalini}
\color{black}

\subsection{Detailed Analysis of Li Insertion, Solvent Effects}

\begin{figure}
\centerline{\hbox{ (a) \epsfxsize=1.50in \epsfbox{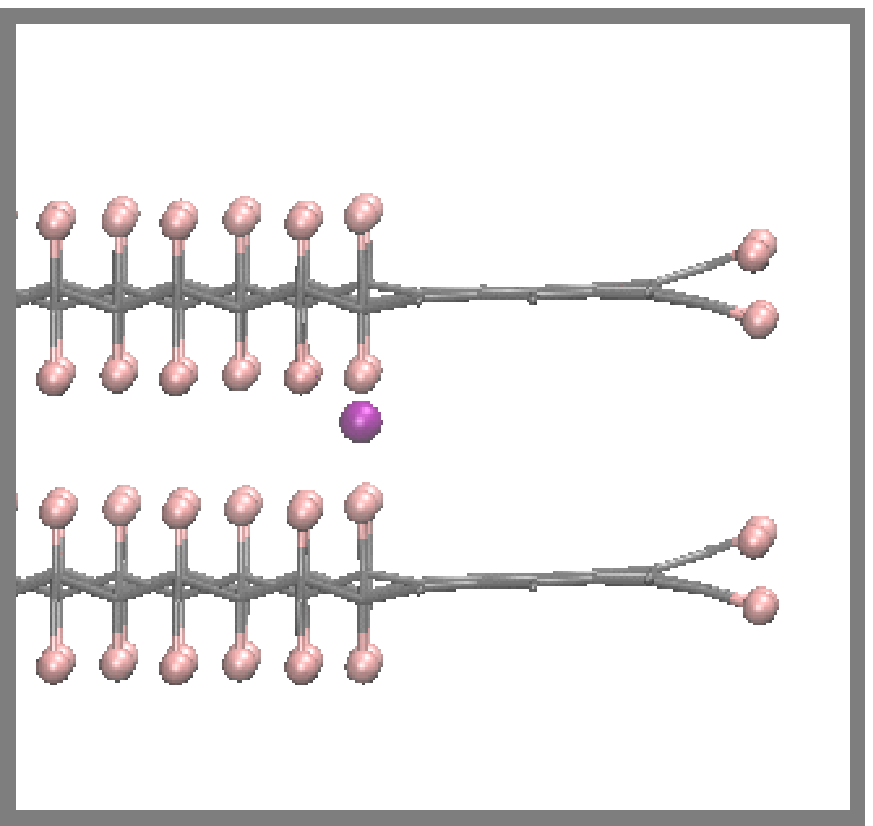} 
		   \epsfxsize=2.80in \epsfbox{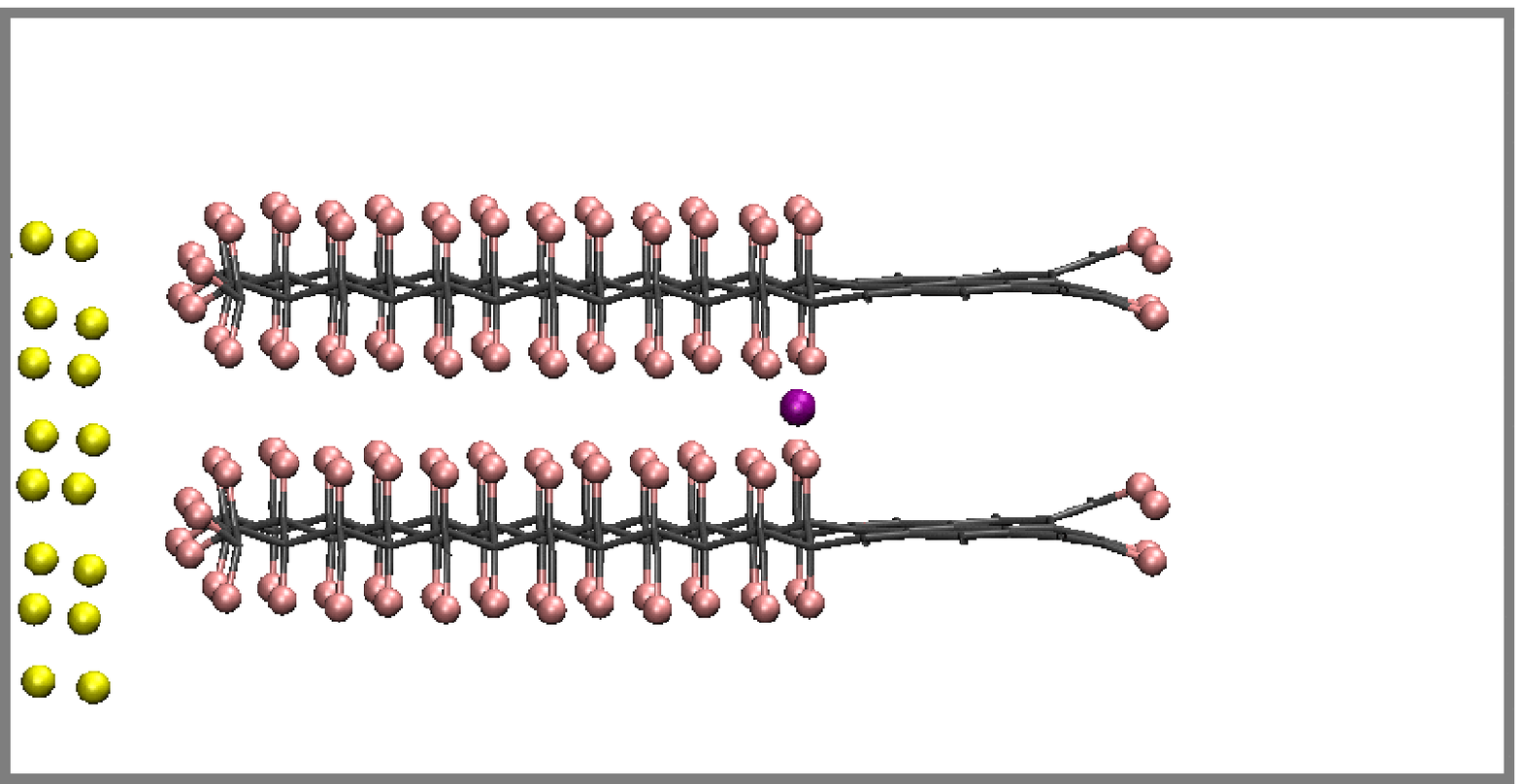} (b)}}
\caption[]
{\label{fig6} \noindent
(a)-(b) Li-insertion at arm-chair edge CF/C interfaces, with and without
an Au current collector (yellow).
}
\end{figure}

Fig.~\ref{fig3}d depicts Li insertion further inside the CF region, away from
the interface.  It is unfavorable in energy by 0.56~eV relative to the
interfacial site (Fig.~\ref{fig3}a), likely due to larger charge separation
between the Li$^+$ and the partially delocalized injected $e^-$.  Thus interior
CF sites would not be occupied by Li unless the discharge occurs at significant
overpotentials.  Thus our proposed interfacial discharge intermediate phase
is small in length scale and may be difficult to detect experimentally.  The
F-removal energy landscape with this zig-zag edge is depicted in
Fig.~\ref{fig4}a.  \color{black} Far from the interface, Li insertion into
CF should revert to the ${\cal V}_i$=0.93~V reported in
Sec.~\ref{edge1}.\color{black}

Fig.~\ref{fig3}e depicts adding a second Li atom at an interfacial site near
the first.  The energy gain in this step corresponds to ${\cal V}_i$=1.45~V,
which is significantly less favorable than the 2.62~V associated with the
first Li.  This finding suggests that the interfacial site is not saturated
with Li.  However, doubling the
simulation cell $y$-dimension (Table~\ref{table1}, lines 6-7, not shown in
figures) is found to reduce ${\cal V}_i$ by only 0.01~V, suggesting that
separation by $\sim$1~nm is the saturation limit for interfacial Li atoms.

After removing one F-atom from Fig.~\ref{fig3}a, the next discharge
event associated with the next Li insertion is 2.72~V (not shown), slightly
higher than the 2.62~V before removing that F-atom.  The C-F bond cleavage
subsequent to this second Li-insertion is also energetically more favorable
than before (-0.27~eV compared with +0.32~eV) and the activation energy is
lower (+0.32~eV compared with 0.53~eV).  The more favorable defluorination
energetics is partly a manifestation of an ``odd-even'' electron spin effect
discussed in the SI (Sec.~S3).  Finally, we consider Li insertion
at the interfacial site after a more substantial defluorination.  Removing
several rows of F atoms to leave another clean-cut CF/C boundary
(Fig.~\ref{fig3}f) yields Li intercalation ${\cal V}_i$=2.56~V at the
interfacial site, which is only slightly
less than the 2.62~V associated with Fig.~\ref{fig3}a.  This conforms with
the Fig.~\ref{fig1}d hypothesis that the defluorination voltage depends on
the local environment, not the global F-content in CF$_x$.  If the F-removal
leaves a jagged boundary between the CF and C regions (not shown),
${\cal V}_i$ is slightly increased, to 3.12-3.15~V depending on the
defluorination extent.  \color{black} We note that row-by-row C-F bond
breaking during discharge is limited by the ``gatekeeper'' site with the
lowest local ${\cal V}_i$, which is in the 2.56-2.62~V range. At or below
such voltages, a new intact row of CF groups can be defluorinated.\color{black}

So far we have not addressed solvent effects.  Fig.~\ref{fig5}a-b depict the
addition of a single ethylene carbonate (EC) or fluoromethane (FM) solvent
molecule to the Li at the 3-coordinated outer binding site (Fig.~\ref{fig3}b);
they dovetail with Eq.~\ref{eq2} proposed in the
literature.\cite{nakajima1999}  EC is a standard battery electrolyte solvent
molecule, similar in structure to PC used in Fig.~\ref{fig2}a-b; the discharge
profile using EC (Fig.~\ref{fig2}c) is similar to that without EC.
FM (featured in Fig.~\ref{fig2}d) is a key component of a recently proposed
liquefied gas electrolyte particularly useful at low temperatures.\cite{fm0,fm}
In Fig.~\ref{fig2}, PC-based and EC-based electrolytes are found to exhibit
discharge plateaus higher in voltage than FM-based electrolyte by 0.1-0.3~V.
Our DFT predictions are that ${\cal V}_i$ is higher for EC than FM, by
2.90~V vs.~2.49~V (Fig.~\ref{fig3}a-b).  This difference is only slightly
higher than the experimental difference, but is significantly less than
the relative gas phase binding energy; we find that an EC binds to
a single Li$^+$ more favorably than FM to Li$^+$, by 1.09~V eV.  \color{black}
Fig.~\ref{fig5}c-f deal with isolated CF$_x$ sheets rather than stacks.  They
are relevant to the exposed outer basal plane surfaces of CF$_x$ stacks and
large curvature, fluorinated nanotubes.\cite{shao2016}  In these cases, no
6-coordinated Li$^+$ binding site exists between two CF$_x$ sheets, and
solvent molecules may be necessary to stabilize Li adsorption.  Here we
postulate that isolated CF$_x$ sheets also undergo row-by-row defluorination
and exhibit sharp interfaces between CF and C regions.  
In Fig.~\ref{fig5}d, we place a Li at the surface site coordinated to
3 CF bonds and an EC molecule on the zig-zag edge.  After optimizating atomic
positions, the Li$^+$ becomes strongly coordinated to two CF groups and
less strongly bound to a third CF (Fig.~\ref{fig5}d).  ${\cal V}_i$ for this
configuration is 2.66~V, less than the 2.90~V for Fig.~\ref{fig5}a but
still in reasonable agreement with experimental measurements (Fig.~\ref{fig2}).
This suggests that our Li$^+$ intercalation mechanism applies to isolated
CF$_x$ sheets, not just stacks.  The corresponding value for FM
(Fig.~\ref{fig5}f) is 2.25~V, less than the 2.49~V for Fig.~\ref{fig5}b.

The ${\cal V}_i$ variation due to EC or FM coordination is 0.41-0.42~V,
qualitatively similar to those found in our experimental measurements
(Fig.~\ref{fig2}), further supporting our computational interpretation of
CF$_x$ discharge behavior.  These ${\cal V}_i$ are also within a few
tenths of a volt of the solvent-free 2.62~V at the 6-coordinated interfacial
site at this F-content (Fig.~\ref{fig3}a); they remain far closer
to experimental discharge voltages than previous DFT calculations.\color{black}

We caution that these equilibrium voltages are calculated using a gas
phase solvent molecule reference.  They ignore solvent-solvent attractive
free energies which may add 0.3-0.5~eV to the cost of each solvent molecule.
\color{black} It is difficult to estimate the EC or FM desolvation free
energy contribution because the electrolytes have mixed solvents and
salts, the DFT/PBE method underestimates dispersion forces between solvent
molecules and between solvent molecules and graphite sheets, and the
entropy contributions involved are difficult to estimate.  However, in
the Fig.~\ref{fig5}d,f configurations, if a liquid solvent were in the
simulation cell, the EC or FM coordinated to Li$^+$ would have been partially
surrounded by solvent molecules.  So these configurations would exhibit far less
desolvation corrections than Fig.~\ref{fig5}a-b, where embedding EC and FM
between graphite sheets would hinder them from interacting with other solvent
molecules.  Despite this difference, the $\Delta {\cal V}_i$ predicted
when using EC (Fig.~\ref{fig5}d) and FM (Fig.~\ref{fig5}f) molecules is almost
the same as that associated with Fig.~\ref{fig5}a-b.  This strongly suggests
that bulk liquid effects do not strongly modify our conclusion about solvent
differences predicted in vacuum.  \color{black}

We stress that the true ${\cal V}_i$ in any electrolyte is bounded from below
by the solvent-free value.  If solvent-coordinated Li$^+$ at the outer binding
site is not more favorable than the solvent-free interfacial site
(Fig.~\ref{fig3}a), the solvent will not contribute to ${\cal V}_i$.  This helps
explain the limited electrolyte-dependence observed during Li/CF$_x$ discharge
(Fig.~\ref{fig2} and Refs.~\onlinecite{solvents,watanabe82,dudney2014}),
compared to the much larger dependence of gas phase binding energies on solvent
molecules.  \color{black} Note also that we choose the solvent models here
to compare with our measurements.  In the future, we will consider solvents
found in commercial Li/CF$_x$ cells like $\gamma$-butyrolactone.\cite{read2009}
\color{black}

\subsection{Arm-Chair Edges}

\begin{figure}
\centerline{\hbox{ (a) \epsfxsize=2.20in \epsfbox{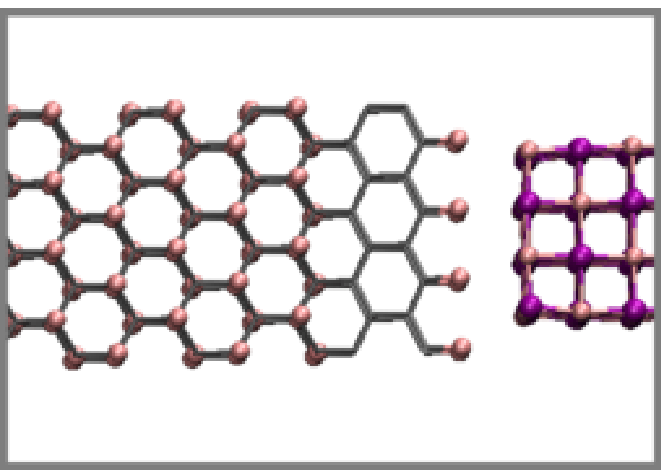} 
		   \epsfxsize=2.20in \epsfbox{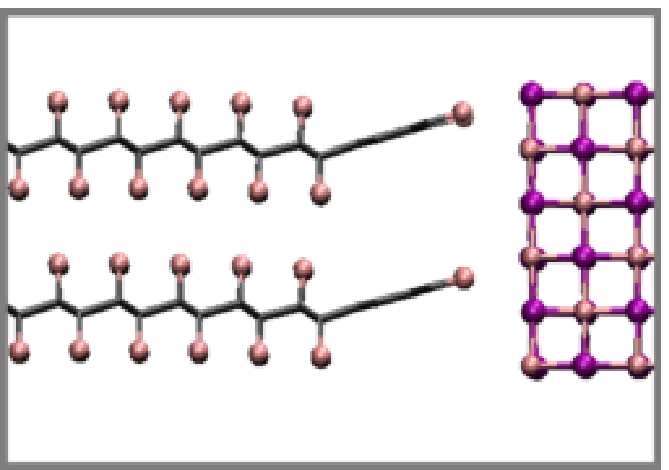} (b)}}
\centerline{\hbox{ (c) \epsfxsize=2.20in \epsfbox{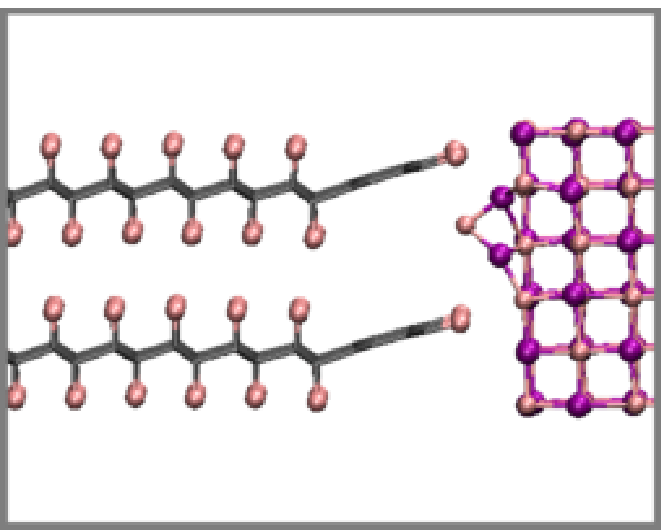} 
		   \epsfxsize=2.20in \epsfbox{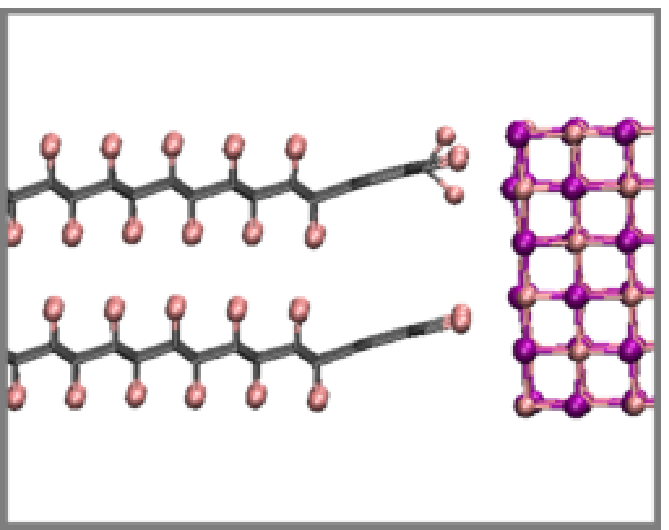} (d)}}
\caption[]
{\label{fig7} \noindent
(a)-(b) Two views of the finite LiF slab at zig-zag surface of CF/C interface;
(c) F$^-$ moved to inner surface; (d) formation of C-F bond.  LiF is three
bilayer ($\sim$12~\AA) thick in the $z$ direction (left-to-right).
}
\end{figure}

As discussed in the SI (Sec.~S4), arm-chair edge CF$_x$ partially defluorinated
in the direction normal to the edge yields insulating carbon because of 
tensile strain.  If the periodic boundary condition is removed in the lateral
direction so the model represents a finite sheet, this carbon will likely
spontaneously develop a curvature, reducing the strain and the electronic band
gap.  However, the CF$_x$/C interface region may still exhibit local
strain, which may be measurable in Raman spectroscopy.\cite{raman1,raman2}
In this work, we focus on periodically replicated arm-chair edges.

Fig.~\ref{fig6}a depicts the insertion of a Li atom into the arm-chair
interfacial site, which is coordinated to 4 C-F groups.  In this configuration,
${\cal V}_i$ is
1.49~V, lower than the zig-zag edge value.  This is consistent with the
fact that it now costs extra energy to deposit an $e^-$ into the conduction
band of the system because of the finite band gap (SI Sec.~S4).  Charge density
changes induced by Li-insertion (Fig.~\ref{fig4}c) show that the excess
$e^-$ still causes somewhat localized changes at the interface, but over a
larger graphite region than zig-zag.

In a working battery cathode, metallic electrode behavior is maintained by
adding conductive carbon additives and depositing the material on to a current
conductor.  CF$_x$ is also converted to conductive carbon upon discharge.  In
Fig.~\ref{fig6}b, we add a Au(100) current collector to the system to
qualitatively mimic a electronically conducting component.  ${\cal V}_i$
is predicted to be 1.81~V, which is 0.32~V more favorable than the model
without Au (Fig.~\ref{fig6}a) but is still significantly lower than the zig-zag
value of 2.62~V, or experimental discharge voltages (Fig.~\ref{fig2}).
Therefore an external current collector does not strongly alter the
equilibrium voltage associated with Li insertion. \color{black} The excess
density plot (Fig.~\ref{fig4}d) shows significant
delocalization of excess $e^-$ over the CF region, but little of the excess
charge is on the Au electrode, \color{black}
likely because the metallic region is too far from the interface to yield
benefits.  We conclude that CF/C interfaces at zig-zag edges are more active
than those at arm-chair edges in terms of CF$_x$ discharge.  In the SI
(Sec.~S5), we consider electronic voltage effects in this system.\cite{pccp}

We have also examined the outer binding site with Li$^+$ coordinated to 3
C-F bonds and an EC molecule (not shown but analogous to Fig.~\ref{fig5}a).
${\cal V}_i$ is 1.96~V in this case, substantially lower than either
measurement with carbonate electrolytes (Fig.~\ref{fig2}a,c) or the
2.92~V zig-zag edge value.  These calculations demonstrate the importance
of a metallic carbon region in CF$_x$ discharge.  

In the SI, we also compare insertion of Na rather than Li into CF$_x$
(Sec.~S8).  The predicted voltages are found to be similar but the C-F
bond-breaking event exhibits a larger barrier and is slower.  We speculate
that multiple Na$^+$ coordinated to one C-F group may reduce the C-F
bond-breaking barrier there and play a role in Na/CF$_x$ discharge.

\subsection{Recharge Calculations}
\label{recharge}

In this section, we return to the Li/CF$_x$ zig-zag edge model
(Fig.~\ref{fig3}a)
and consider Li/CF$_x$ recharge.  We add a LiF slab outside the
partially defluorinated CF$_x$ edge (Fig.~\ref{fig7}a, Table~\ref{table2}).
The slab is periodically replicated in the direction perpendicular to the
CF$_x$ sheets, but is finite in extent in the other lateral direction
(Fig.~\ref{fig7}a-b).  This baseline model system mimics discharged LiF
residing on the defluorinated CF$_x$ edge.

Next, a F$^-$ anion is moved from the outer LiF surface to its inner surface
closest to the CF$_x$ edge, where it is now coordinated to two Li$^+$ ions
(Fig.~\ref{fig7}c).  This configuration mimics voltage-driven F$^-$ diffusion
through the LiF discharge product towards the defluorinated carbon, where
it ultimately reconstitutes a C-F bond.
The energy cost associated with creating the defect (Fig.~\ref{fig7}c)
from the defect-free sheet (Fig.~\ref{fig7}a-b) is 2.44~eV in this 
12.2$\times$10.4$\times$44.0~\AA$^3$ simulation cell.  This periodically
replicated cell is charge neutral, but exhibits a signficant dipole moment
which can lead to electrostatic energy artifacts.\cite{silicatm}  Using
larger, 1$\times$2$\times$1 and 2$\times$2$\times$1 surface supercells, the
energy costs are $\Delta E$=2.14~eV and~2.21~eV respectively, suggesting that
this energy has converged to within $\sim$0.1~eV.  During recharge, this energy
cost needs to be provided by changing the applied electronic voltage 
${\cal V}_e$.\cite{pccp}  A more complete discussion of ${\cal V}_e$ is given
in the SI (Sec.~S5).
Fig.~\ref{fig7}d depicts a configuration where the F$^-$ on the inner LiF
surface is further transferred to the carbon region to form a C-F bond.
$\Delta E$=-0.26~eV relative to Fig.~\ref{fig7}c, and the barrier is a
small $\Delta E^*$=0.19~eV.  This suggests that reformation of C-F
bonds from nearby undercoordinated F$^-$ anions can readily occur;
formation of this bond is not the prohibitive rate-limiting step.

To show that a voltage change can cause F$^-$ diffusion from the LiF
outer surface to its inner surface prior to C-F bond formation, we
create an artificial electric field by removing a Li$^+$ from the
outer interface of the LiF slab.  In terms of electronic voltage,\cite{pccp}
the Li$^+$ vacancy causes the Fig.~\ref{fig7}a-b ${\cal V}_e$ to increase
from 4.02~V to 6.46~V.  This latter ${\cal V}_e$ is artificially high,
partly because of the lack of a liquid electrolyte beyond the 
outside surface of the LiF slab.  However, ${\cal V}_e$ includes contributions
from many interfaces, including the LiF/vacuum and C/LiF interfaces which
should be similar before and after Li$^+$ vacancy formation.  It is the 
electric field across the LiF slab, reflected in the voltage difference
between the two cases ($\Delta {\cal V}_e$=+2.44~V), that is relevant to our
calculation.  
%The potential drop associated with
%$\Delta {\cal V}_e$occurs entirely inside the LiF slab, not in the electric
%double layer associated with the electrolyte.  

Upon creating the Li$^+$
vacancy on the outer LiF surface, the Fig.~\ref{fig7}c configuration is
no longer metasable, unlike the case without the Li$^+$ vacancy.  The
$\Delta E$ for moving a F$^-$ from the outer LiF surface (Fig.~\ref{fig7}d)
to the inner surface and reforming a C-F bond decreases from an
endothermic +2.19~eV to an exothermic (favorable) -0.93~eV in the
1$\times$1$\times$1 surface cell. The corresponding $\Delta E$ for the
2$\times$2$\times$1 surface cell is even more favorable at -1.45~eV.
These $\Delta E$'s strongly suggest that refluorination, and recharging,
are energetically and kinetically viable near the outer defluorinated
CF$_x$ surface in contact with LiF discharge products if a sufficient
electric field is applied.

Note that we have focused on reforming a C-F bond at the outer
defluorinated CF$_x$ surface, not the CF/C interface which is the
focus of the discharge calculations (Fig.~\ref{fig3}).  Therefore we
assume recharge occurs ``from outside in.''   The reason is that we
assume F$^-$ anions are poorly solvated by organic solvent molecules.
This suggests that moving F$^-$ to the CF/C interface may entail a
large barrier.  (In contrast, Li$^+$ is much better solvated by organic
solvents and can be readily transported to the CF/C interface via
Li$^+$(solvent molecule)$_n$ complexes and can initiate discharge there.)
If this hypothesis is correct, CF$_x$ charge and discharge will follow
different pathways.   This may be an important consideration when
designing practical rechargeable Li/CF$_x$ cathodes.

\section{Conclusions}

The existence of a CF$_x$ discharge ``intermediate phase'' has been proposed
in the literature to explain the discharge behavior of Li/CF$_x$ batteries,
but its identity has so far not been elucidated.  In this work, we hypothesize
an intermediate phase associated with a CF$_x$ edge-propagation Li-insertion
mechanism.  Our DFT calculations show that Li intercalation at the zig-zag
edge boundary between fluorinated (i.e., CF) and defluorinated (carbon) regions
exhibits an equilibrium voltage about 2.6~V vs.~Li$^+$/Li(s).  The predicted
voltage range is in good agreement with experimental CF$_x$ electrochemical
discharge voltage profiles, and is only 0.4-0.5~V lower than GITT measurements.
Li-insertion is favorable
there because of partial separation of Li$^+$ and $e^-$ charges at the
interface, in accordance with ``interfacial charge storage''
paradigm.\cite{maier2014}  Na/CF$_x$ is predicted to exhibit similar
discharge voltages (SI).  Our proposed CF$_x$ intermediate phase has a spatial
extent limited to the CF/C interface thickness.  These predictions should
assist renewed experimental attempt to locate the CF$_x$ intermediate phase.
Our predicted voltages do not involve C-F bond breaking or LiF nucleation
energetics.  

This proposed mechanism is not inconsistent with the existence of a thin,
several-atom-thick LiF layer that may reside between CF$_x$ sheets,
swelling the cathode material in the process.  Indeed, our hypothesis may
provide a viable pathway to realize a CF$_x$:Li$_y$ intermediate phase.  We
also show that post Li-insertion CF$_x$ discharge kinetics, especially C-F
bond breaking, are sufficiently fast in Li/CF$_x$ batteries when a C-F
bond is polarized by the inserted Li$^+$ cation nearby.  The proposed
discharge mechanism helps explain why solvent dependence on discharge
voltage should be small.  Our predicted variation in voltage plateau
values as the electrolyte varies is in qualitative agreement with 
our discharge measurements in organic carbonate and liquified gas electrolytes.
Finally, we propose that Li/CF$_x$ recharge may proceed via a pathway distinct
from the one encountered during discharge.

Our computational work does not invoke phase diagram calculations, and 
therefore does not assume that CF$_x$ and LiF are in equilibrium.  Since
Li/CF$_x$ batteries are not rechargeable so far, the equilibrium assumption
appears untenable.  Thus our work provides a novel, general framework
for understanding and modeling the discharge behavior of new types of
primary batteries.  For example, in recent work on SF$_6$-based primary
batteries, the equilibrium assumption underlying traditional phase
diagram approach also seems inapplicable.\cite{gallant}  \color{black}
For future work, electric field and electronic voltage effects need to be
addressed more explicitly by including liquid electrolytes in more rigorous
AIMD simulations.  CF$_x$ heterogeneity\cite{nmr1,nmr2,nmr3} will also
be considered to further improve our discharge voltage predictions.\color{black}

\section*{Supporting Information for Publication}
Supporting information is available free of charge on the ACS Publications
website at DOI:xxx: experimental details; F-vacancy diffusion barriers
in CF$_x$; odd-even energetic effects on defluorination; local electronic
densities of state; electronic voltage discussions; entropy considerations;
short-circuit AIMD simulations; Na/CF$_x$ predictions.

\section*{Acknowledgement}
 
We thank Brennan Walder, Todd Alam, Jessica Rimsza, and Hayley Hirsh for useful
input and discussions.  This project was supported by Sandia Laboratory Directed
Research and Development (LDRD) project 218253.  Sandia National Laboratories
is a multi-mission laboratory managed and operated by National Technology
and Engineering Solutions of Sandia, LLC., a wholly owned subsidiary of
Honeywell International, Inc., for the U.S. Department of Energy's National
Nuclear Security Administration under contract DE-NA-0003525.
This paper describes objective technical results
and analysis. Any subjective views or opinions that might be expressed in
the paper do not necessarily represent the views of the U.S.~Department
of Energy or the United States Government.

\section*{References}

\newpage
\begin{figure}
\centerline{\hbox{ \epsfxsize=3.00in \epsfbox{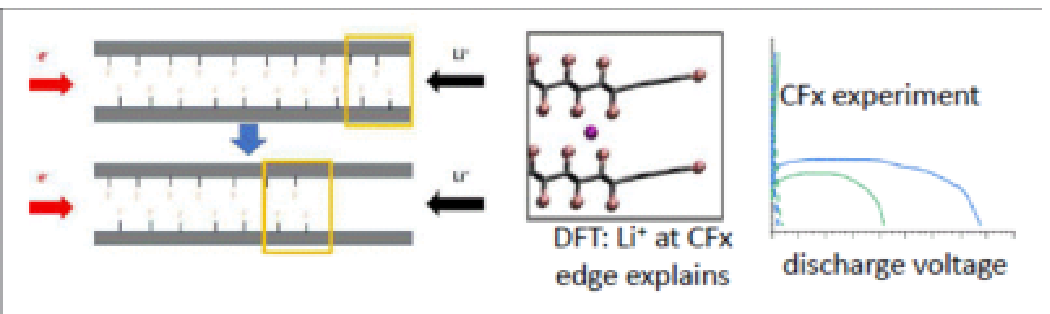} }}
\caption[]
{Table of content Graphic.}
\end{figure}

\end{document}